\documentclass[letterpaper,twocolumn,10pt]{article}
\usepackage{usenix,epsfig,endnotes}

\usepackage{comment}
\usepackage{multirow}
\usepackage{textcomp}
\usepackage{color}
\usepackage[normalem]{ulem}
\usepackage{setspace}
\usepackage{enumerate}
\usepackage{multirow}
\usepackage{algorithm}
\usepackage{algpseudocode}
\usepackage{amsmath}

\usepackage{colortbl}
\usepackage{color}
\usepackage{tabularx}
\usepackage{caption}
\usepackage{makecell}
\usepackage{array}
\usepackage[normalem]{ulem}
\usepackage{titlecaps}

\newcommand{\cb}{\textcolor{blue}}

\usepackage{titling}
\newcommand{\squishlist}{
  \begin{list}{$\bullet$}
  { \setlength{\itemsep}{0pt}      \setlength{\parsep}{-0pt}
    \setlength{\topsep}{4pt}       \setlength{\partopsep}{0pt}
    \setlength{\listparindent}{-2pt}
    \setlength{\itemindent}{-5pt}
    \setlength{\leftmargin}{1em} \setlength{\labelwidth}{0em}
    \setlength{\labelsep}{0.5em} } }

\newcommand{\squishend}{
    \end{list}  }

\newcommand{\nonl}{\renewcommand{\nl}{\let\nl\oldnl}}

\begin{document}

\date{}

\title{\Large \bf A Robust Fault-Tolerant and Scalable Cluster-wide Deduplication for Shared-Nothing Storage Systems 
}


\author{
	{\rm Awais Khan, Chang-Gyu Lee, Prince Hamandawana\thanks{Mr. Prince is currently affiliated with Ajou University, Suwon, Republic of Korea.} , Sungyong Park, Youngjae Kim}\\
	{\rm Sogang University, Seoul, Republic of Korea}\\
} 
\maketitle



\setstretch{0.944}
 
 

\begin{abstract}
	\vspace{-0in}
	
	Deduplication has been largely employed in distributed storage systems to improve 
	space efficiency. 
	Traditional deduplication 
	research 
	ignores the design specifications of shared-nothing distributed storage systems such as no central metadata bottleneck, scalability, and storage rebalancing.  
	Further, 
	deduplication introduces 
	transactional changes,
	which are prone to errors in the event of a system failure, 
	resulting in 
	inconsistencies in data and deduplication metadata.
	In this paper, we propose a robust, fault-tolerant and scalable cluster-wide deduplication 
	that can 
	eliminate 
	duplicate copies across the cluster. 
	We design a distributed deduplication metadata shard which guarantees 
	performance scalability while preserving the design constraints of shared-nothing storage systems. 
	The placement of chunks and deduplication metadata is made cluster-wide based on the content fingerprint of chunks. 
	To ensure transactional consistency and garbage identification, we employ a flag-based asynchronous consistency mechanism. We implement the proposed deduplication 
	on Ceph. The evaluation shows high disk-space savings with minimal performance degradation as well as high robustness in the event of sudden server failure. 

\end{abstract}


\section{Introduction}
\label{intro}
\vspace{-0.15in}


The shared-nothing storage systems (SN-SS) accommodate a large number of storage servers for 
high performance, 
scalability, 
availability, and fault-tolerance~\cite{ceph, glusterfs, echdfs}.
SN-SS such as GlusterFS~\cite{glusterfs} and Ceph~\cite{ceph} is widely employed in cloud storage due to multiple 
properties:
(i) 
it contains no central metadata bottleneck, so it is highly scalable,
(ii) storage servers are independent where a single storage server failure cannot crash the whole cluster, and 
(iii) 
it allows dynamic changes in the cluster, such as addition or removal of storage servers and can relocate 
objects in the cluster to balance storage 
utilization across the storage servers. 





Deduplication (dedup) techniques are employed in cloud storage systems to increase storage efficiency. 
There exist 
several
studies on cluster-wide deduplication (dedup)~\cite{extreme, venti, probability, datadomain, dede, scalableinline, silo, hydrastor, probability, exact, primarysystor}. However, 
direct adoption of such dedup techniques
on the SN-SS
violates the basic design constraints of 
SN-SS. 
For example, 
A central dedup server approach using a single deduplication metadata management server~\cite{exact, venti, extreme, datadomain} 
not only violate shared-nothing 
properties of SN-SS but also 
limit the scalability. 
On the other hand, 
a decentralized approach to distributing deduplication metadata management across multiple servers~\cite{scalableinline, hydrastor, EMC, dede, silo, probability, boafft, marknsweep} require additional hardware and software resource cost for multiple deduplication servers. 
In order to reduce such additional cost, simple DB-sharding approach that embeds 
the DB-shard of the whole dedup metadata database on each storage server has been proposed~\cite{exact}.
However, 
this DB-sharding approach to SN-SS 
suffers from inherited problems, i.e., to identify a duplicate chunk, the fingerprint lookup must be broadcasted to all DB-shards in the cluster.

\begin{figure}[!t]
	\begin{center}
		\begin{tabular}{@{}c@{}c@{}}
			\includegraphics[width=0.5\textwidth]{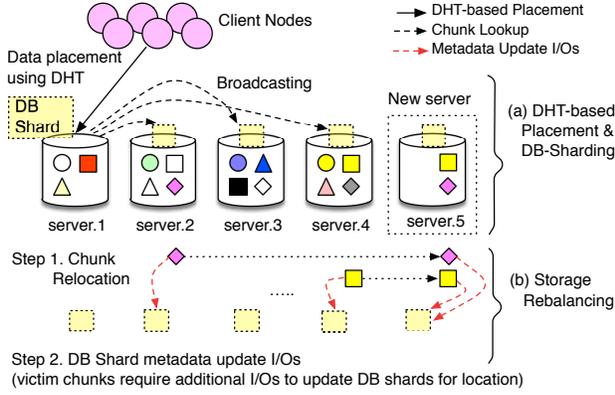} \\
		\end{tabular}
		\vspace{-0.1in}
		\caption{\small{(a) Traditional distributed DB sharding approach and (b) storage rebalancing issues in SN-SS such as Ceph~\cite{ceph} and GlusterFS~\cite{glusterfs}.
				Specifically, (b) illustrates the chunk relocation when a new server is added to the cluster. 
		}}
		\label{fig:dbshard}
	\end{center}
	\vspace{-0.3in}
\end{figure}

Another challenging issue is deeply related to
storage rebalancing. 
In SN-SS, the storage rebalancing is triggered whenever a change such as adding or deleting a storage server in cluster occurs. This rebalancing shuffles the chunks across the storage servers to evenly balance the space utilization across the storage servers in the cluster. 
In this case, deduplication metadata must be updated for the new location of the chunk in the cluster.
However, this rebalancing incurs high metadata I/Os to renew DB-shards on each storage server. 
Figure~\ref{fig:dbshard}(a)(b) illustrate these problems.
Deduplication also requires transactional level changes, where a complete object-based transaction is split into multiple small fixed or variable chunk-based transactions~\cite{exact}. 
These changes, if not implemented carefully, can cause inconsistent data and deduplication metadata in the cluster in an event of storage server failures. 
A recent study to address the consistency of reference counts is to use soft-update style metadata in a single disk-based file system~\cite{order}. 
However, it is not directly applicable to distributed nature of SN-SS, where parallel I/Os are responsible to distribute chunks. Additionally, transaction ordering and delay operation require additional checkpointing and journaling overhead which is contrary to deduplication, i.e., space savings. The undo~\& redo logging can be employed but due to additional space overhead and fingerprint lookup latency, the storage cost increases. 
Another effect of transaction failures in deduplication storage systems is 
garbage chunks 
of failed transactions. 










To address the above-mentioned challenges in SN-SS, 
we propose to build a scalable and consistent cluster-wide deduplication framework for SN-SS.
In particular, we use chunk's content fingerprint to avoid lookup broadcast issue in DB-shard and employ a tagged consistency approach to ensure the validity of deduplication metadata. This paper has following specific contributions:

\vspace{-0.03in}
\squishlist
\item
We employ database partitioning to handle deduplication metadata in a decentralized manner and preserve the shared-nothing property of SN-SS. We use the content-based fingerprint to distribute and locate the chunks in the cluster. Even if chunks are shuffled across the storage servers in the cluster, content-fingerprint is able to determine the exact location of the storage server responsible for storing the object and chunks. 

\item
We aim to design asynchronous tagged consistency which ensures correct status of the transaction and deduplication metadata. Moreover, our partitioned deduplication metadata and tagged consistency aid in identifying garbage chunks and require no additional monitoring and journaling.  

\item 
We design and implement the proposed data deduplication in Ceph, a scale-out shared-nothing storage system~\cite{ceph} and evaluate our proposed ideas in real testbeds. 
\squishend
\vspace{-0.15in}

\section{Cluster-wide Data Deduplication}
\label{design}
\vspace{-0.1in}


\subsection{Architecture Overview}
\label{overview}
\vspace{-0.1in}

\begin{figure}[!t]
	\begin{center}
		\begin{tabular}{@{}c@{}c@{}}
			\includegraphics[width=0.47\textwidth]{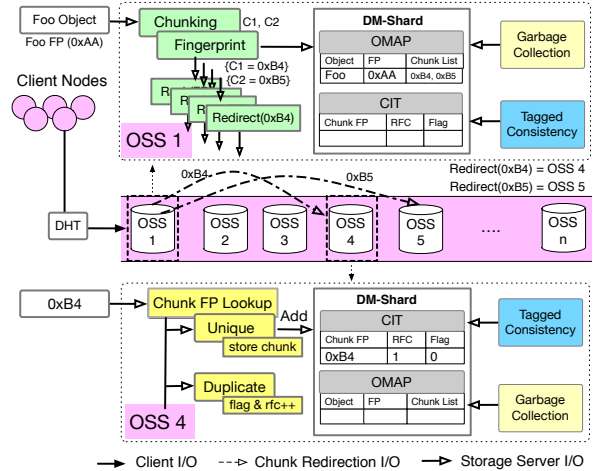} & \\
		\end{tabular}
		\vspace{-0.15in}
		\caption{\small{
				Cluster-wide  deduplication based on DB-sharding and content-fingerprint based placement in SN-SS. 
		}}
		\label{fig:overview}
	\end{center}
	\vspace{-0.35in}
\end{figure}

The proposed cluster-wide deduplication is built on a shared-nothing distributed storage system. 
Figure~\ref{fig:overview} shows the architecture design of cluster-wide deduplication. 
Logically, the SN-SS is composed of clients, storage servers and no additional metadata servers and employs distributed-hash table (DHT) for data placement~\cite{ceph, glusterfs}. 

The client performs object name hashing and locates the storage server to write or read objects in the cluster. 
Each storage server performs deduplication and stores data and metadata.
When storage server receives a write request (OSS 1 in Figure~\ref{fig:overview}), it is responsible for splitting the object into small fixed-size data chunks, computes the fingerprint for each chunk's content. Then, it redirects the data chunk 
to storage server based on the computed fingerprint (OSS 4 in Figure~\ref{fig:overview}).


This fingerprint-based redirection frees from keeping the location of each data chunk in the storage system. 
At this point, the storage server builds a mapping of the object and its data chunks' fingerprints 
in DM-Shard (Deduplication Metadata Shard) 
as shown in Figure~\ref{fig:overview} (OSS 1). We explain the DM-Shard in Section~\ref{dbshard} in detail.
The redirected chunks received on other storage servers (OSS 4 in Figure~\ref{fig:overview}) are treated in the following manner; 
The chunk fingerprint lookup is made in CIT (Chunk Information Table) of DM-Shard, which is responsible for maintaining the fingerprint of data chunk, reference count and commit flag. 
The reference count of a fingerprint shows the degree of references linked to it and commit flag is a tag to ensure the validity of the chunk (tagged consistency), i.e., whether the fingerprint is pointing to valid stored content in the storage server or content is missing from the storage server.
If chunk fingerprint exists and commit flag is valid, then  
the reference count (RFC in CIT) increment is granted. Whereas, the non-existence of fingerprint is treated as a unique chunk. The data chunk is stored in the storage server and CIT entry is updated accordingly (OSS 4). This process is iterated for all the data chunks in parallel. When all the chunks are stored, then Object-Map (OMAP) entry is created (OSS 1) which defines the object layout such as, name, fingerprint and chunk list of the object. The write operation finishes, when all the data chunks, OMAP and CIT data structures are created.




The tagged consistency
guarantees the validity and correctness of all the CIT entries and data chunks in storage without additional logging and journaling. The DM-Shard and tagged consistency 
together assist in identifying the garbages and orphan data chunks, i.e., remains of partially failed transactions. The chunk fingerprints with an invalid flag (Flag in CIT) are interpreted as garbage data chunks and collected periodically. 

\vspace{-0.15in}

\begin{figure*}[!t]
	\centering
	\begin{tabular}{@{}l@{}r@{}}
		\includegraphics[width=0.97\textwidth]{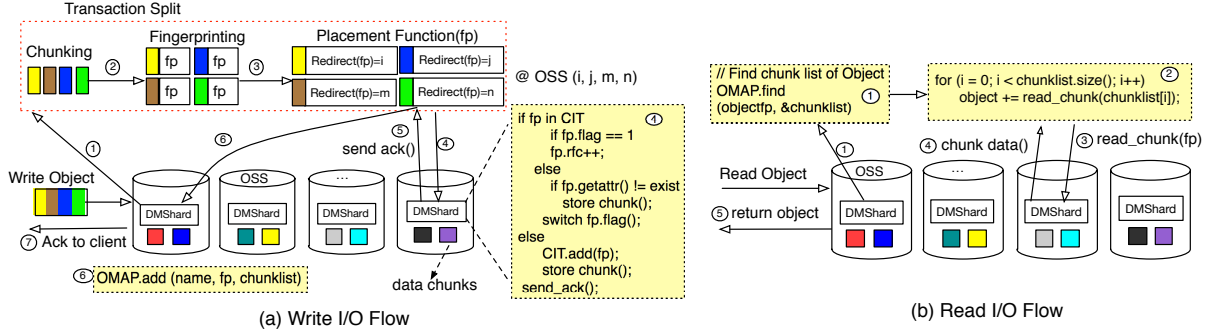} \\
	\end{tabular}
	\vspace{-0.15in}
	\caption{\small A complete write and read I/O transaction in cluster-wide data deduplication system. }
	\vspace{-0.2in}
	\label{fig:dedupflow}
\end{figure*}

\subsection{Deduplication  Metadata Shard}
\label{dbshard}
\vspace{-0.1in}

We build a Deduplication Metadata Shard (DM-Shard)  as
shown in Figure~\ref{fig:overview} to effectively manage deduplication metadata. 
The design decision to use distributed DM-Shard is to comply scalable and shared-nothing property of SN-SS. Every storage server in the cluster hosts a DM-Shard holding all the persistent data structures such as object layout information 
and data chunk fingerprint. 
Each shard keeps the unique information 
of objects and data chunks in a separate data structure, i.e., Object Map (OMAP) and Chunk Information Table (CIT). 



\squishlist
\item
{\em Object Map (OMAP):} 
OMAP maintains the complete layout and reconstruction logic of an object, i.e., object name, object fingerprint, and list of data chunks. The OMAP data structure is shown in Figure~\ref{fig:overview}. 
In DHT-based storage systems, an object is identified by hashing the object name, and if we do not maintain the hash of object, we cannot reconstruct the original object because we need all the chunks' fingerprint created from this object.
OMAP assists in read operations, where object fingerprint is given to lookup chunks belonging to a specific object.

\item
{\em Chunk Information Table (CIT):} 
CIT maintains the performance-sensitive deduplication metadata. It includes data chunk fingerprint, 
reference count and commit flag. All the lookup and reference update operations are possible via this data structure. 
\squishend

The advantage to 
keep different data structures is 
manifold: i) to provide effective execution of fingerprint operation, i.e., lookup, increment/decrement, ii) reduced congestion on a single data structure when multiple I/Os access the data structure, iii) to avoid data chunk fingerprint lookup in case of the read request.

Both OMAP and CIT data structures are updated synchronously during a write operation to avoid concurrent lookups of identical fingerprints, which can result in storage inefficiency. We describe complete read and write I/O transaction with usage of OMAP and CIT in Figure~\ref{fig:dedupflow}.
For deduplication metadata replication and fault-tolerance, we rely on SN-SS because we store our DM-Shard in the storage server and is replicated like a normal object.


\vspace{-0.15in}
\subsection{Chunk Relocation and I/O Routing}
\label{shuffling}
\vspace{-0.1in}

SN-SS such as Ceph~\cite{ceph} and Gluster~\cite{glusterfs} distribute objects in a storage-balanced fashion. For instance, Ceph uses CRUSH algorithm~\cite{crush} to fairly distribute the storage load across the storage servers, when the cluster topology changes, e.g., a new storage server is added, removed or disk failure occurred.
The objects are relocated across the storage servers in order to balance the storage load in the cluster as shown in Figure~\ref{fig:dbshard}(b). 
This object and chunk relocation process is neglected in 
all previous deduplication studies~\cite{scalableinline, hydrastor, exact, boafft}. In previous studies, the location of object and data chunks 
is stored along with metadata, i.e., \textit{data chunk 1A is stored on server x and data chunk 1B is stored on server y}. 
This type of deduplication metadata management suffers when chunks are relocated in the cluster because object and chunk location is lost. One solution can be; to transform
current self-balancing mechanism to update the deduplication metadata while relocating the objects and chunks, 
but it entails complex implementation and a high number of I/Os for every object and chunk relocation to update the deduplication metadata. 

To determine the exact location of the data chunk and related DM-Shard across the cluster, we use the data chunk fingerprint. 
The fingerprint can be obtained in two ways: i) to generate the fingerprint directly from the data chunk contents (write request approach), and ii) to obtain the data chunk fingerprint from OMAP using object name or object fingerprint (read request). The computed-fingerprint tells the storage server location responsible for storing the actual data chunk and the metadata shard (CIT). 
This content-based placement relieves us from i) complicated location management for each data chunk, ii) modifications in existing self-balancing mechanism, and iii) frequent deduplication metadata updates. Another gain of this content-based placement is that we do not require to broadcast I/Os to all storage servers for fingerprint lookup, instead we send a single lookup I/O to only a single storage server.



\vspace{-0.15in}
\subsection{Asynchronous Tagged Consistency}
\label{tagedconsistency}
\vspace{-0.08in}

The deduplication metadata inconsistencies in distributed storage systems lead to data authenticity and integrity issues. For example, if an object transaction is split into multiple chunk-based transactions, and one of the small transactions fails. Then, in such case, the whole object transaction fails and two problems are likely to happen. First, an invalid reference fingerprint in DB-shard and second, garbage chunks left of the failed transaction. Worst of all, new incoming duplicate fingerprint increments the invalid reference entry, causing serious metadata inconsistency. Due to transactional modifications, a complicated transaction and rollback logic is required to make reference count consistent~\cite{marknsweep, physicalgarbage}.

To address 
such consistency concern, we add a commit flag
to each data chunk entry which specifies the consistency state of the chunk, i.e., 0 or 1.
The flag with 0 is invalid chunk (missing from storage) and 1 is valid chunk (available in storage). 
A simple approach is to add commit flag with object or chunk data structure and update the commit flag at transaction completion time. 
However, this simple approach requires transaction lock and updates the flag synchronously which affects the scalability of the system.  To bypass such transaction lock, we propose an asynchronous thread-based consistency manager which runs on every storage server. All the incoming write I/Os registers to consistency manager. Once the I/O transaction completes, the consistency manager asynchronously updates the flag managed in CIT (Section~\ref{dbshard}). If a crash occurs in the middle of a transaction when data chunk is stored and commit flag is not updated, then, in such case, the chunk will be marked as garbage due to invalid commit flag value because transaction partially failed. 
We explain the tagged consistency using two use cases.






{\em Unique Write:}
In this case, the object splits into multiple small chunks and stores the chunk on different storage servers based on data chunk fingerprint. Each fingerprint in CIT holds an invalid flag by default, i.e., 0. The consistency manager is notified of the received write operation.  Once the I/O finish, the flag is switched from invalid (0) to valid (1) asynchronously.


{\em Duplicate Write:} In duplicate write case, whenever a duplicate fingerprint wants to increment the reference count in CIT, it needs to check the flag as shown in Figure~\ref{fig:dedupflow}. The fingerprint entries with a valid flag allow the reference count increment or decrement operations. If the flag is invalid and reference update is required. Then, the data chunk is required to perform an additional consistency check, to ensure the existence of data chunk in the storage server. We manage consistency check by simply getting data chunk attributes from the storage server just like a stat call in the file system. If the data chunk exists, we switch the flag to valid and conduct the reference operations. Otherwise, we first store the actual data chunk contents and then, switch the flag. This consistency check enables the presence of actual data and can repair the missing data chunks.


To claim free space consumed by garbage data chunks, we design and implement a garbage collection thread. The thread periodically collects the data chunk fingerprints with an invalid commit flag in CIT. It keeps the fingerprints for a pre-defined threshold. Once the threshold expires, the thread cross-match the collected FPs to CIT. This cross-matching is required to assess any change, in particular to 
invalid fingerprints. If there is no change, then fingerprints along with data chunks are removed from the storage system. We do not use any additional journaling because it requires additional disk space. We claim that the proposed asynchronous consistency manager ensures the data and metadata accuracy even in case of failures and prevent the deduplication storage system from inconsistencies. 





\section{Evaluation}
\label{eval}
\vspace{-0.15in}




{\bfseries {Implementation:}}
We implement the proposed cluster-wide deduplication framework in Ceph v10.2.3. 
The DM-Shard, 
consistency manager and garbage collector are embedded in each OSD (Object Storage Daemon). 
We use the SHA-1 algorithm to generate 
a data chunk fingerprint and pass the fingerprint to the CRUSH algorithm~\cite{crush} 
to distribute the data chunks in the Ceph storage cluster. 
We use SQLite~\cite{sqlite} as backend storage for DM-Shard.

{\bfseries {Testbed:}}
We configured the Ceph storage cluster with four Object Storage Servers (OSS) 
equipped with two SSDs acting as Object Storage Daemons (OSD). The details of testbed are listed in Table~\ref{tab:testbed}.
We used the FIO~\cite{fio} benchmark with librbd/krbd support for evaluation by varying deduplication ratio and and number of client threads with a 500GB workload. 
We compare the proposed cluster-wide deduplication technique with Baseline Ceph without deduplication and Ceph with a central server deduplication. We drop cache after every experiment and report the average of 5 iterations for each experiment. 


\begin{table}[h]
	\centering 
	\vspace{-0.1in}
	{\scriptsize %
		\begin{tabular}{|c|c|}
			\hline
			\rowcolor[gray]{.9}   \multicolumn{2}{|c|}{OSS (x4) \& Monitors (x3)}  \\
			\hline
			Processor & \makecell{Intel(R) Xeon(R) CPU E5-2640 v4 \\@ 2.40GHz(10 cores)} \\
			\hline
			Memory & 32GB \\
			\hline
			Network & 10Gbps \\
			\hline
			OS & \makecell{CentOS 7.3.1611} \\ 
			\hline
			Storage & \makecell{Samsung SSD 850 PRO 256GB x 2 Per OSS} \\
			\hline
			\hline
		\end{tabular}
	}
	\vspace{-0.1in}
	\caption{{\small Testbed setup.}}
	\label{tab:testbed}
	\vspace{-0.2in}
\end{table}

\begin{figure}[!t]
	\centering
	\begin{tabular}{@{}c@{}c@{}}
		\includegraphics[width=0.25\textwidth]{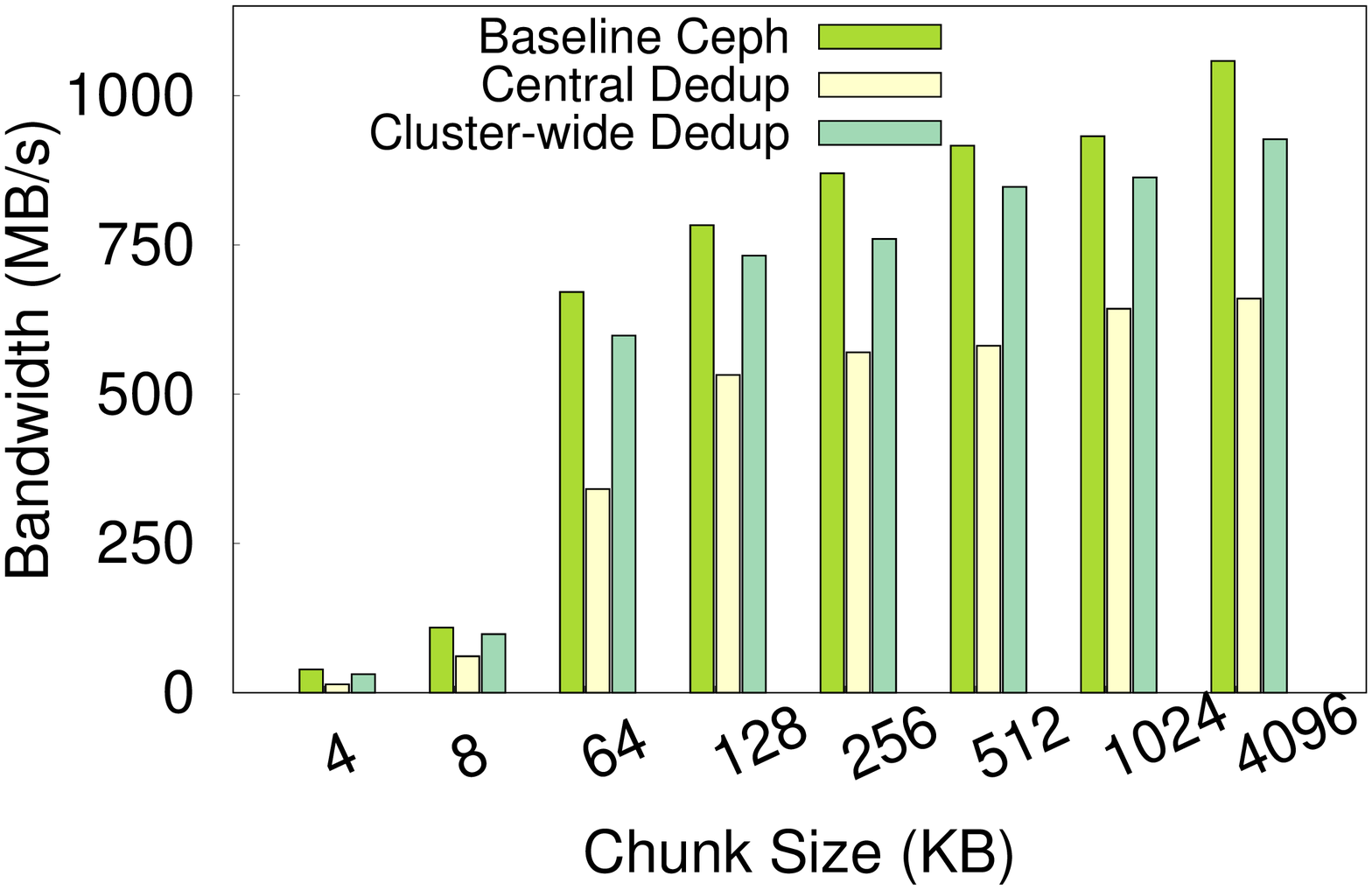} &
		\includegraphics[width=0.25\textwidth]{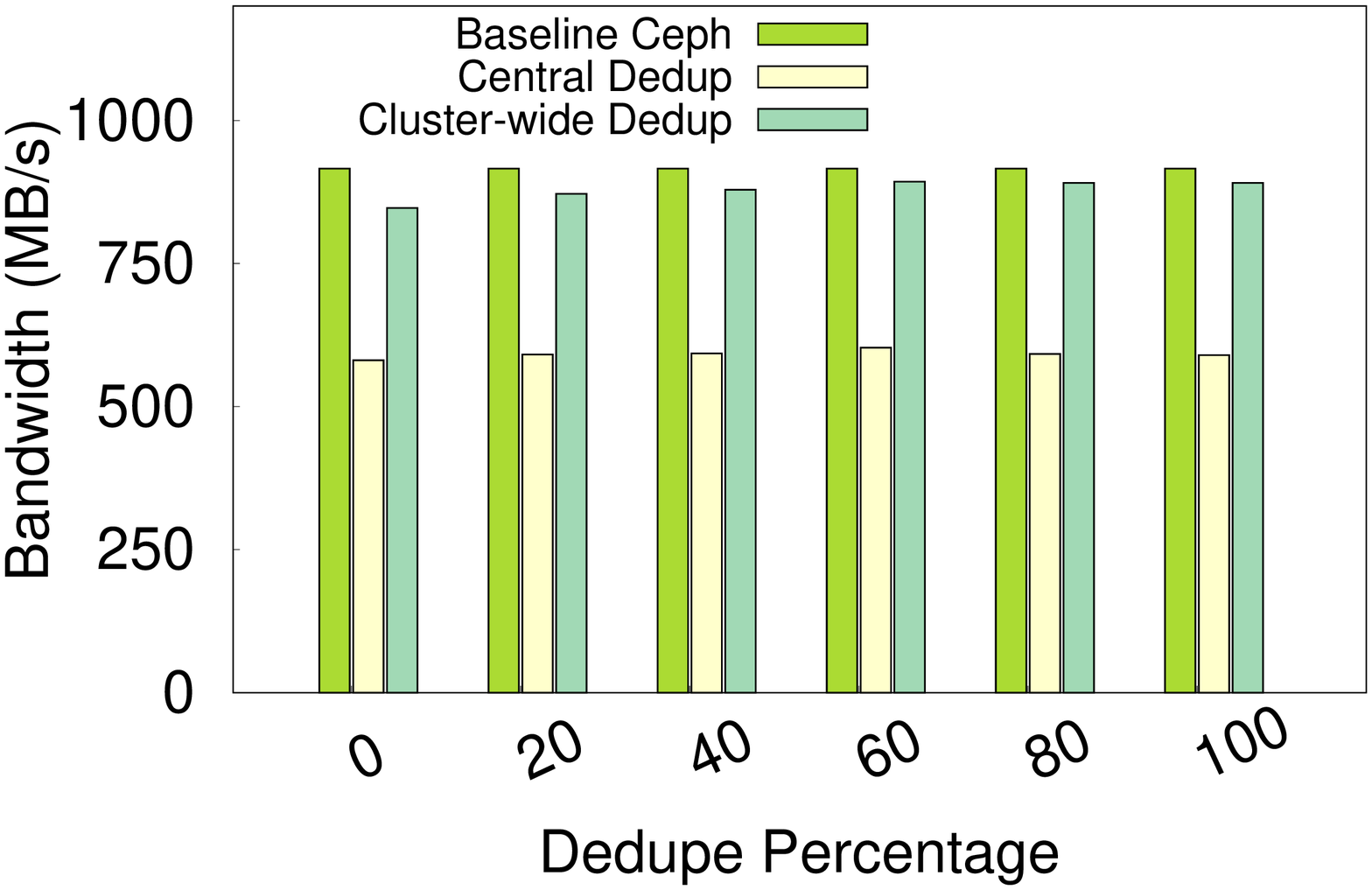}  \\	
		\vspace{-0.1in}	
		(a) {\scriptsize Performance Analysis} &
		(b) {\scriptsize Deduplication Ratio} \\	 
	\end{tabular}
	\vspace{-0.05in}
	\caption{\small Performance analysis.}
	\vspace{-0.2in}
	\label{fig:perf}
\end{figure}

{\bfseries{Performance Analysis:}}
To analyze the performance penalty incurred by the proposed cluster-wide dedup, we use synthetic datasets generated via FIO~\cite{fio}. To clearly observe the performance overhead, we set the deduplication percentage to 0\% and use 8 client threads in FIO benchmark. Figure~\ref{fig:perf} (a) shows the bandwidth of all three approaches. 
Our proposed partitioned metadata scales as much as baseline Ceph with respect to the increased chunk size. However, there is a certain performance overhead which is mainly derived from fingerprint computation and network overhead for small chunk-sizes. The fingerprint overhead can be further minimized by employing hardware-accelerator such as GPU for parallel fingerprint computation.

Next, we discuss the performance of cluster-wide dedup with respect to deduplication ratio as shown in Figure~\ref{fig:perf} (b). We set the chunk size to 512KB and use 8 client threads to 
compare the central and
cluster-wide dedup approaches. 
We observe both central deduplication and cluster-wide dedup approaches show limited performances to certain thresholds regardless of deduplication rate. 
However, we see 
the cluster-wide dedup performance is twice that of central dedup.
This improvement is basically due to scalable and distributed deduplication metadata management, which reduces the metadata I/O contention. 
We do not observe notable performance improvement with cluster-wide dedup when dedup ratio varies because small data chunk I/Os are still directed over the network which are too small to show improvement if not stored on the storage server. 


\begin{figure}[!t]
	\centering
	\begin{tabular}{@{}c@{}c@{}c@{}c@{}c@{}}
		\includegraphics[width=0.25\textwidth]{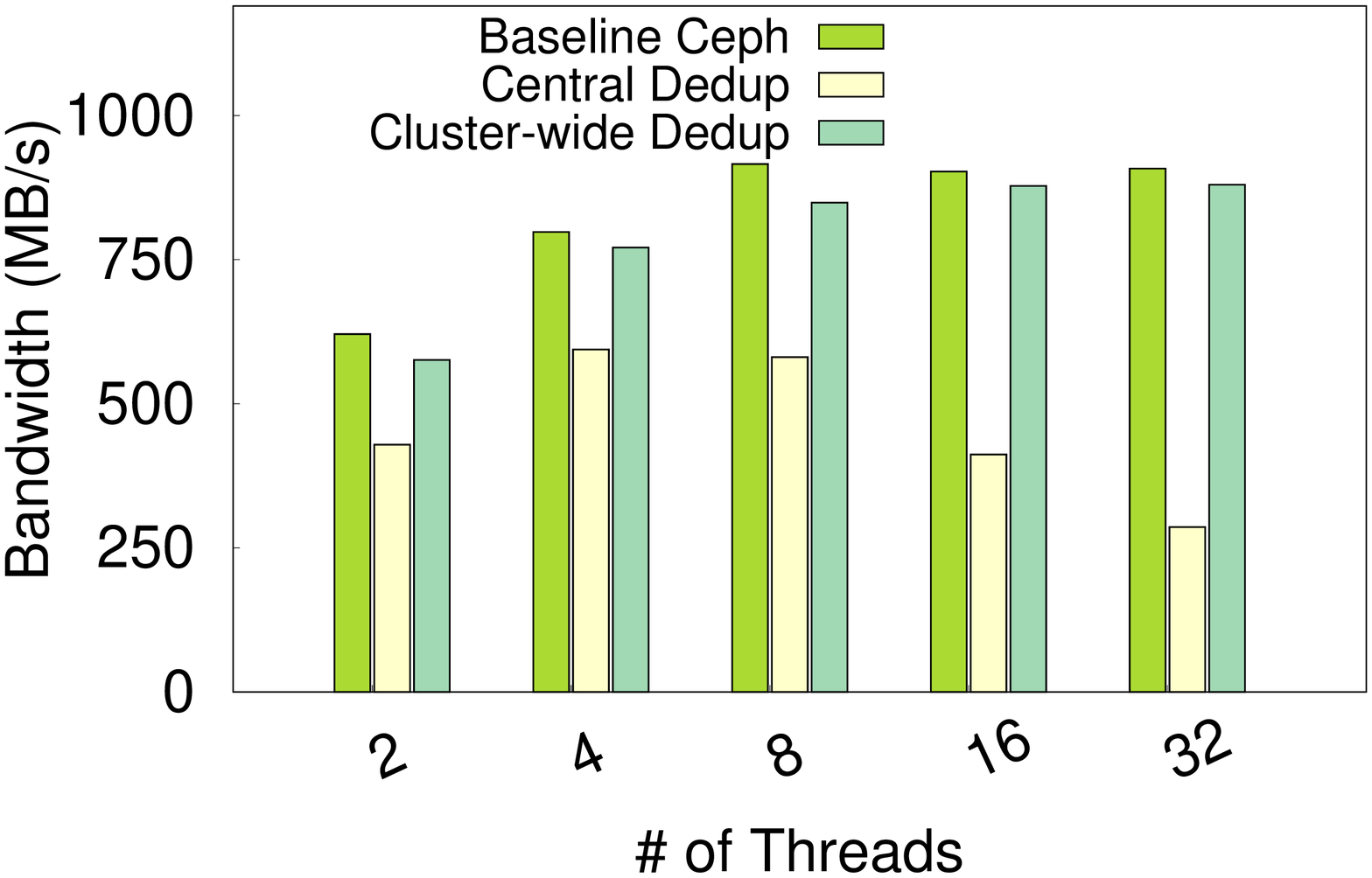} &
		\includegraphics[width=0.25\textwidth]{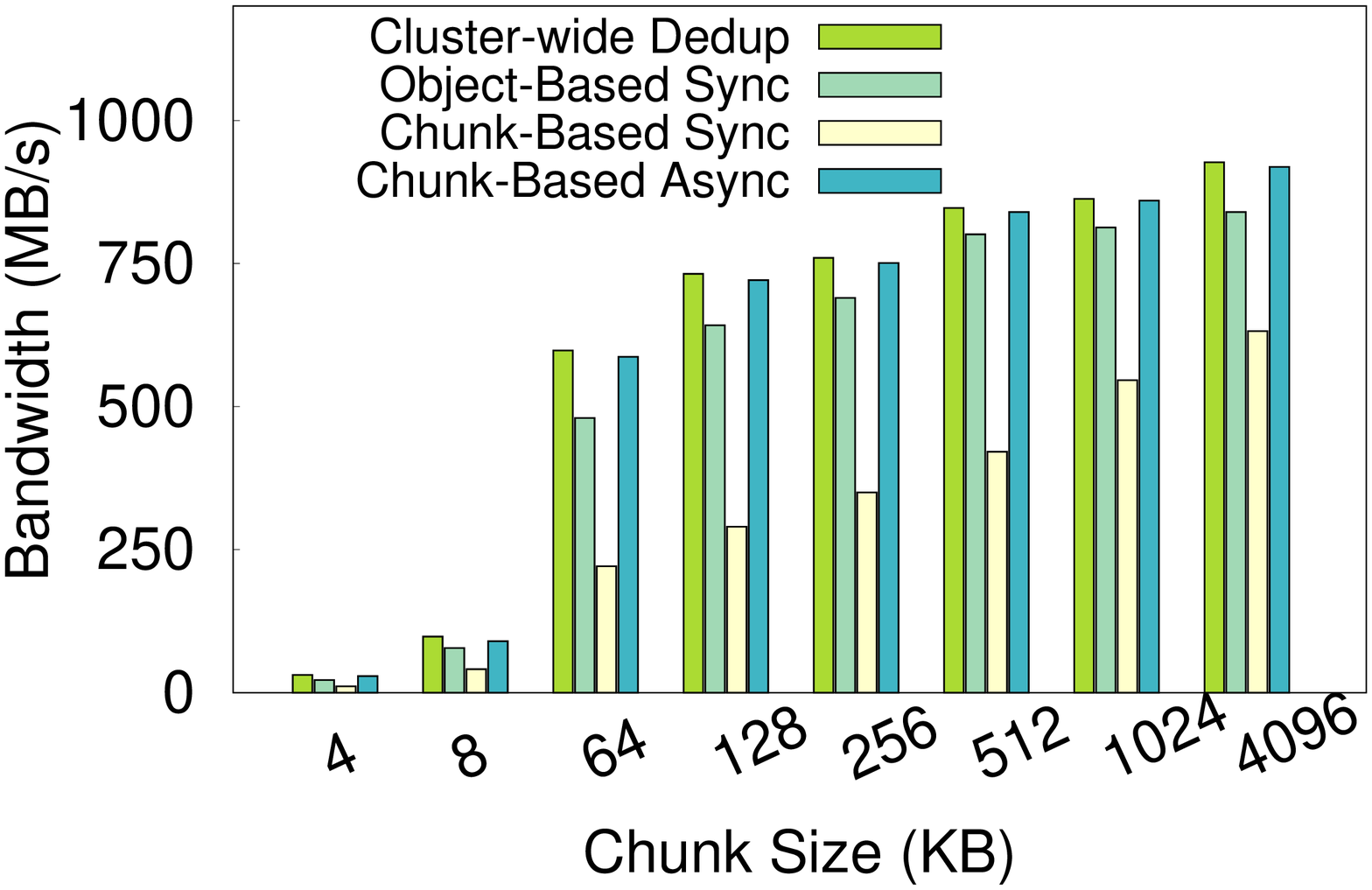} & \\
		\vspace{-0.1in}	
		(a)  {\scriptsize{Scalability with Multiple Clients}} &
		(b) {\scriptsize Asynchronous Tagged Consistency} & \\		 
	\end{tabular}
	\vspace{-0.05in}
	\caption{\small Scalability and consistency analysis.
	}
	\vspace{-0.25in}
	\label{fig:expr}
\end{figure}

{\bfseries{Scalability Analysis:}}
To test the scalability, we use multiple client threads in FIO~\cite{fio}. 
In Figure~\ref{fig:expr}~(a), we tend to show the impact of I/O contention created by multiple clients. We set the chunk size 512KB to benefit the counter approach, i.e., central dedup because single deduplication metadata DB becomes a bottleneck due to increased number of concurrent I/Os.
Figure~\ref{fig:expr}~(a) shows that, when the number of client threads is less, the cluster-wide dedup performance is high compared to central dedup even when there is no contention. This is because central dedup server is responsible for all the chunking and fingerprinting overhead. However, with the increased number of client threads, central dedup further degrades the performance. 
It becomes worse when the number of client threads is 32, the central dedup bandwidth degrades 
to 200MB. Whereas, 
our cluster-wide deduplication approach shows scalability and improves the bandwidth with increasing number of client threads because CRUSH~\cite{crush} distributes the data chunks uniformly in a load-aware fashion to object storage servers and DM-Shard is distributed across all the object storage servers which overcome the possible chances of dedup metadata contention.

{\bfseries{Asynchronous Tagged Consistency:}}
In chunk-based consistency, the flag 
is managed for each data chunk fingerprint, whereas in object-based consistency, the flag is stored at object granularity.  Figure~\ref{fig:expr}~(b) shows the bandwidth of different variant when employed. We see that, when chunk size is small, the performance is poor in both chunk and object-based consistency compared to asynchronous tagged consistency. However, when we increase the chunk size, the performance improves. The chunk-based consistency shows high performance overhead as compared to others. It is due to additional serialized number of I/Os required to switch flags. Whereas, in object-based consistency shows fair performance because only a single I/O is required to switch the flag but still degrades the performance more than 15\% compared to baseline cluster-wide deduplication. On the other hand, the asynchronous tagged consistency incurs negligible overhead compared to chunk and object-based consistency overhead. Because both chunk and object consistency approaches introduce a transaction lock which increases the I/O latency, whereas our approach switches the commit flags asynchronously without acquiring any transaction lock, hence no overhead is incurred.   


\begin{table}[!t]
	\centering
	\scriptsize
	\begin{tabular}{|l|l|l|l|l|l|l|l|l|}
		\hline
		\multirow{2}{*}{\textbf{Deduplication}} & 
		\multicolumn{2}{c}{}{}{{}}{\textbf{\# of Disks}} &
		\multicolumn{2}{c|}{} \\ \cline{2-5} 
		& 1 & 2 & 4 & 8 \\ 
		\hline
		{Cluster-wide Dedup Approach} & 85 & 85 & 85 & 85  \\
		\hline
		{Disk-based Dedup Approach}  & 85 & 77 & 65 & 61 \\
		\hline
	\end{tabular}
	\vspace{-0.1in}
	\caption{\small Deduplication space savings in percentage. }
	\label{tab:space}
	\vspace{-0.25in}
\end{table}

{\bfseries{Storage Efficiency:}}
We conduct this experiment to show the storage space efficiency of proposed cluster-wide dedup compared to local disk-based deduplication. To enable disk-based dedup, we configure Ceph cluster with BtrFS~\cite{btrfs} as backend disk file system with deduplication enabled. We use 100\% deduplication ratio and report the results in Table~\ref{tab:space}. We observe that disk-based dedup storage efficiency decreases with increasing number of disks. It is because disks are not aware of each other and cannot identify the duplicates stored on other disks. 
Whereas, 
cluster-wide dedup storage efficiency remains high irrespective of number of disks.





\vspace{-0.1in} 
\section{Concluding Remarks}
\label{conc}
\vspace{-0.1in}

This paper presents a robust fault-tolerant, cluster-wide deduplication framework for shared-nothing storage systems. We design and implement a distributed deduplication metadata shard approach that uses the content hash of chunks
to avoid I/O 
broadcasting and dynamic object relocation problems. 
We also propose a tagged consistency approach which can 
recover reference errors and lost data chunks in case of sudden storage server failures. 
We implement the proposed ideas 
on Ceph. 
The evaluation shows that our proposed approaches support high scalability with minima performance overhead and high robust fault tolerance. 
Our future work is to 
minimize the fingerprint overhead and evaluation on a large-scale testbed with realistic datasets.


\vspace{-0.15in}
 
{\footnotesize \bibliographystyle{acm}
\bibliography{refs.bib}}

\end{document}